\documentstyle[preprint,aps,epsf,floats]{revtex}

\begin{document}
\tighten
\def\si{{}^1\kern-.14em S_0}
\def\siii{{}^3\kern-.14em S_1}
\def\diii{{}^3\kern-.14em D_1}
\def\pone{{}^3\kern-.14em P_1}
\def\pzero{{}^3\kern-.14em P_0}
\def\ptwo{{}^3\kern-.14em P_2}
\newcommand{\gsim}{\raisebox{-0.7ex}{$\stackrel{\textstyle >}{\sim}$ }}
\newcommand{\lsim}{\raisebox{-0.7ex}{$\stackrel{\textstyle <}{\sim}$ }}
\def\pislash{ {\pi\hskip-0.6em /} }
\def\pislashsmall{ {\pi\hskip-0.375em /} }
\def\pslash{p\hskip-0.45em /}
\def\nopi{ {\rm EFT}(\pislash) }
\def\nopit{ {\rm dEFT}(\pislash) }
\def\Ltwo{ {^\pislashsmall \hskip -0.2em L_2 }}
\def\Lone{ {^\pislashsmall \hskip -0.2em L_1 }}
\def\CQuad{ {^\pislashsmall \hskip -0.2em C_{\cal Q} }}
\def\Czero{ {^\pislashsmall \hskip -0.2em C_{0}^{(\siii)} }}
\def\Czeromone{ {^\pislashsmall \hskip -0.2em C_{0,-1}^{(\siii)} }}
\def\Czerozero{ {^\pislashsmall \hskip -0.2em C_{0,0}^{(\siii)} }}
\def\Czeroone{ {^\pislashsmall \hskip -0.2em C_{0,1}^{(\siii)} }}
\def\Ctwo{ {^\pislashsmall \hskip -0.2em C_{2}^{(\siii)} }}
\def\Ctwomtwo{ {^\pislashsmall \hskip -0.2em C_{2,-2}^{(\siii)} }}
\def\Ctwomone{ {^\pislashsmall \hskip -0.2em C_{2,-1}^{(\siii)} }}
\def\Cfour{ {^\pislashsmall \hskip -0.2em C_{4}^{(\siii)} }}
\def\CSDzero{ {^\pislashsmall \hskip -0.2em C_0^{(sd)} }}
\def\CSDtwotwotwo{ {^\pislashsmall \hskip -0.2em C_{2,-2}^{(sd)} }}
\def\CSDzeromone{ {^\pislashsmall \hskip -0.2em C_{0,-1}^{(sd)} }}
\def\CSDzerozero{ {^\pislashsmall \hskip -0.2em C_{0,0}^{(sd)} }}
\def\CSDtwoone{ {^\pislashsmall \hskip -0.2em \tilde C_2^{(sd)} }}
\def\CSDtwotwo{ {^\pislashsmall \hskip -0.2em C_2^{(sd)} }}
\def\CSDzerotwo{ {^\pislashsmall \hskip -0.2em C_{0,0}^{(sd)} }}
\def\LX{ {^\pislashsmall \hskip -0.2em L_X }}
\def\CSDfour{ {^\pislashsmall \hskip -0.2em C_4^{(sd)} }}
\def\CSDfourt{ {^\pislashsmall \hskip -0.2em \tilde C_4^{(sd)} }}
\def\CSDfourtt{ {^\pislashsmall \hskip -0.2em {\tilde{\tilde C}}_4^{(sd)} }}
\def\etasd{\eta_{sd} }
\def\ZCzeromone{ 
{_z \hskip -0.4em {^\pislashsmall \hskip -0.2em C_{0,-1}^{(\siii)} }}}
\def\ZCzerozero{ 
{_z \hskip -0.4em {^\pislashsmall \hskip -0.2em C_{0,0}^{(\siii)} }}}
\def\ZCzeroone{ 
{_z \hskip -0.4em {^\pislashsmall \hskip -0.2em C_{0,1}^{(\siii)} }}}
\def\ZCtwomtwo{ 
{_z \hskip -0.4em {^\pislashsmall \hskip -0.2em C_{2,-2}^{(\siii)} }}}
\def\ZCtwomone{ 
{_z \hskip -0.4em {^\pislashsmall \hskip -0.2em C_{2,-1}^{(\siii)} }}}
\def\ZCfourmthree{ 
{_z \hskip -0.4em {^\pislashsmall \hskip -0.2em C_{4,-3}^{(\siii)} }}}
\def\rCzeromone{ 
{_\rho \hskip -0.4em {^\pislashsmall \hskip -0.2em C_{0,-1}^{(\siii)} }}}
\def\rCzerozero{ 
{_\rho \hskip -0.4em {^\pislashsmall \hskip -0.2em C_{0,0}^{(\siii)} }}}
\def\rCzeroone{ 
{_\rho \hskip -0.4em {^\pislashsmall \hskip -0.2em C_{0,1}^{(\siii)} }}}
\def\rCtwomtwo{ 
{_\rho \hskip -0.4em {^\pislashsmall \hskip -0.2em C_{2,-2}^{(\siii)} }}}
\def\rCtwomone{ 
{_\rho \hskip -0.4em {^\pislashsmall \hskip -0.2em C_{2,-1}^{(\siii)} }}}
\def\rCfourmthree{ 
{_\rho \hskip -0.4em {^\pislashsmall \hskip -0.2em C_{4,-3}^{(\siii)} }}}
\def\CPzero{ {^\pislashsmall \hskip -0.2em C^{(\pzero)}_2  }}
\def\CPone{ {^\pislashsmall \hskip -0.2em C^{(\pone)}_2  }}
\def\CPtwo{ {^\pislashsmall \hskip -0.2em C^{(\ptwo)}_2  }}

\def\Journal#1#2#3#4{{#1} {\bf #2}, #3 (#4)}

\def\NCA{\em Nuovo Cimento}
\def\NIM{\em Nucl. Instrum. Methods}
\def\NIMA{{\em Nucl. Instrum. Methods} A}
\def\NPB{{\em Nucl. Phys.} B}
\def\NPA{{\em Nucl. Phys.} A}
\def\NP{{\em Nucl. Phys.} }
\def\PLB{{\em Phys. Lett.} B}
\def\PRL{\em Phys. Rev. Lett.}
\def\PRD{{\em Phys. Rev.} D}
\def\PRC{{\em Phys. Rev.} C}
\def\PRA{{\em Phys. Rev.} A}
\def\PR{{\em Phys. Rev.} }
\def\ZPC{{\em Z. Phys.} C}
\def\SJP{{\em Sov. Phys. JETP}}
\def\SJNP{{\em Sov. Phys. Nucl. Phys.}}

\def\FBS{{\em Few Body Systems Suppl.}}
\def\IJMP{{\em Int. J. Mod. Phys.} A}
\def\UJP{{\em Ukr. J. of Phys.}}
\def\CJP{{\em Can. J. Phys.}}
\def\SCI{{\em Science} }
\def\AST{{\em Astrophys. Jour.} }
\def\tran{dibaryon}
\def\trans{dibaryons}
\def\Tran{Dibaryon}
\def\Trans{Dibaryons}
\def\TRANS{DIBARYONS}
\def\yt{y}

\preprint{\vbox{
\hbox{ NT@UW-00-028}
}}
\bigskip
\bigskip

\title{Rearranging Pionless Effective Field Theory}

\author{{\bf Silas R. Beane}$^a$ and {\bf Martin J. Savage}$^{a,b}$}
\address{$^a$ Department of Physics, University of Washington, \\
Seattle, WA 98195. }
\address{$^b$ Jefferson Laboratory, 12000 Jefferson Avenue,\\
Newport News, VA 23606.}
\maketitle

\begin{abstract}
We point out a redundancy in the operator structure of the pionless
effective field theory, $\nopi$, which dramatically simplifies
computations. This redundancy is best exploited by using
\tran{} fields as fundamental degrees of freedom. 
In turn, this suggests a new power counting scheme which sums range
corrections to all orders.  We explore this method with a few simple
observables: the deuteron charge form factor, $np\rightarrow d\gamma$,
and Compton scattering from the deuteron.  Unlike $\nopi$, the higher
dimension operators involving electroweak gauge fields are not
renormalized by the s-wave strong interactions, and therefore do not
scale with inverse powers of the renormalization scale.  Thus, naive
dimensional analysis of these operators is sufficient to estimate
their contribution to a given process.
\end{abstract}

\vskip 2in

\vfill\eject


\section{Introduction}

An effective field theory ($\nopi$) has been constructed that allows
for precise calculations of very low-energy nuclear processes.  In
this energy regime, all multi-nucleon interactions are described by
local operators.  The relative size of the contribution of an operator
to a particular observable can be estimated by the power-counting
rules developed in~\cite{Lu95,vK97,Ka98B,Ka98A,Ka99C,vK98,Ch99}.  An array
of phenomena have been studied with $\nopi$, such as the radiative
capture $np\rightarrow d\gamma$ relevant to Big-Bang
nucleosynthesis~\cite{CS99,Ru99}, elastic and inelastic $\nu d$
scattering~\cite{BC00,BC00B} relevant to the ongoing efforts to
measure the solar-neutrino flux and the cross section for
$pp\rightarrow de\nu$~\cite{Ko99A,Ko99B,Ko99C,Ko00}, the
electromagnetic form factors of the deuteron~\cite{Ch99,Ch99B}, and
other processes involving electroweak gauge fields
(for a detailed review of this subject see Ref.~\cite{Ioffe}).

While calculation of such processes to high precision is
straightforward, the number of Feynman diagrams involved can become
quite large.  It was shown that determining the coefficients in the
Lagrange density by reproducing the deuteron binding energy, $B$, and
the residue of the deuteron pole, $Z_d$, at next-to-leading order
(NLO)~\cite{Ph99}, instead of reproducing the scattering length,
$a^{(\siii)}$ and effective range, $r^{(\siii)}$ at NLO dramatically
improves the convergence of $\nopi$. This is 
because the effective range in both s-wave channels is somewhat
larger than one would naively guess, the expansion parameter 
in the $\siii$ channel  being
$\gamma r^{(\siii)}\sim 0.4$, where $\gamma^2=M_N B$ is the deuteron
binding momentum.  

One of the simplifications we suggest in this work is that
$r^{(\siii)}$ should be taken to be of order $1/Q$ in the power
counting, a suggestion first made in Ref.~\cite{Ka99}.  This naturally
leads to the use of \tran{} fields, as first discussed in
Ref.~\cite{Ka97} and profitably exploited in
Ref.~\cite{bosons,harald,haraldb}.  In this formulation of $\nopi$,
operators are ordered according to powers of the total center-of-mass
energy. We will see that this simplification is a consequence of a
redundancy in the operator structure of $\nopi$. Moreover, in this
modified power counting, \tran{} operators representing
higher-dimensional nucleon contact interactions are not renormalized
by the s-wave interactions as in standard $\nopi$~\cite{Ch99} and as a
consequence are ordered according to naive dimensional analysis.
These observations lead to a remarkable simplification in the
computation of higher order effects.

\section{An Operator Redundancy}

To begin with, we make the following general observation that
simplifies computations in $\nopi$.  In higher order calculations,
Galilean-invariant operators appear involving a large number of
derivatives acting on the incoming and outgoing nucleon fields.  By
integrating by parts and then using the equations of motion for the
nucleon field the number of higher order operators is greatly reduced,
and calculations are simplified further by utilizing operators
involving time-derivatives and not just spatial derivatives.  To show
this we consider the interaction Lagrange density that contributes to
NN scattering in the $\siii$-channel at next-to-next-to-leading order
(N$^2$LO) in the EFT expansion,
\begin{eqnarray}
{\cal L}^{N^2 LO} & = & 
- C_4 
\left( N^T P^j (\overleftarrow\nabla - \overrightarrow\nabla)^2 
N \right)^\dagger
\left( N^T P^j(\overleftarrow\nabla - \overrightarrow\nabla)^2 
N \right)
\nonumber\\
& & - {1\over 2}
\tilde C_4
\left[
\left( N^T P^j (\overleftarrow\nabla - \overrightarrow\nabla)^4 
N \right)^\dagger
N^T P^j N
\ +\ 
{\rm h.c.} \right]
\ \ \ ,
\label{eq:cfour}
\end{eqnarray}
where the spin-isospin projector for the $\siii$ channel is
\begin{eqnarray}
P^i & \equiv &  {1\over \sqrt{8}} \sigma_2\sigma^i\  \tau_2
\ \ \ , 
\ \ \  {\rm Tr} \left[ P^{i\dagger} P^j \right]\  =\ {1\over 2} \delta^{ij}
\ \ \ .
\label{eq:progT}
\end{eqnarray}
It is straightforward to show that, by using the equations of motion
for the nucleon field and integrating by parts where necessary,
\begin{eqnarray}
-\theta 
\ N^T P^j (\overleftarrow\nabla - \overrightarrow\nabla)^2 N
& = & 
4 M_N \ \theta\ 
\left[ i \partial_0 + {\nabla^2\over 4 M_N}\right]  N^T P^j N
\equiv 
4 M_N \ \theta\ {\cal O}_E N^T P^j N
\ \ \ ,
\end{eqnarray}
where $\theta$ is some operator, and we have not shown terms that
are total derivatives.
The operator ${\cal O}_E$ when acting on the two-nucleon operator
simply yields the non-relativistic center-of-mass energy.
Therefore, the Lagrange density in eq.~(\ref{eq:cfour}) 
becomes
\begin{eqnarray}
{\cal L}^{N^2 LO} & = & 
- (4 M_N)^2\ 
\left( C_4 + \tilde C_4 \right)
\left[ {\cal O}_E N^T P^j N \right]^\dagger
{\cal O}_E N^T P^j N
\ \ \ ,
\label{eq:cfourE}
\end{eqnarray}
and involves only one operator.  It is clear how to generalize this
result to higher orders in the $\nopi$ expansion, and in fact, there
is only one new operator for each higher order in the energy
expansion.  This is completely consistent with the Effective Range
(ER) expansion~\cite{Be49,BL50}, and with the previous works on
$\nopi$~\cite{Ch99} where it was shown that NN scattering constrains
only the combination $C_4+\tilde C_4$. An all-orders argument for NN
scattering has been given previously~\cite{birse}. The one place where
these operator relations require a little thought is where electroweak
currents are inserted. Actually, the above argument holds when
derivatives are replaced by covariant derivatives. Previous $\nopi$
calculations have shown that the $C_4$ and $\tilde C_4$ operators
contribute differently to electroweak observables.  However, it is
always a linear combination of these operators and a gauge invariant
higher dimension operator that appears, and it is this combination
that is fit to data.  Thus using the operator structure of
eq.~(\ref{eq:cfourE}) will lead to different RG evolution of gauge
invariant higher dimension operators compared to that resulting from
eq.~(\ref{eq:cfour}), but physics will be left unchanged.  This
generalizes the observation made in ref.~\cite{Fu00a,Fu00b}.


\section{Enter \TRANS}

If we take seriously the notion that both $a$ and $r$ are of order
$Q^{-1}$, then we can describe NN scattering with \tran{} fields in
both the $\si$ and $\siii$ channels.  The Lagrange density describing
the dynamics of the \tran{} $t^j$ in the $\siii$ channel is
\begin{eqnarray}
{\cal L}_t & = & 
N^\dagger \left[ i\partial_0 + {\nabla^2\over 2 M_N}\right] N
\ -\ 
t^{\dagger}_a \left[  i\partial_0 + {\nabla^2\over 4 M_N} - \Delta \right] t^a
\ -\  \yt \left[\  t^\dagger_j \ N^T P^j N\ +\ {\rm h.c.} \right]
\ \ \ ,
\label{eq:trandef}
\end{eqnarray}
where $y$ is the coupling between nucleons in the $\siii$ channel and the 
$\siii$-\tran.
%
\begin{figure}[!ht]
\centerline{{\epsfxsize=6in \epsfbox{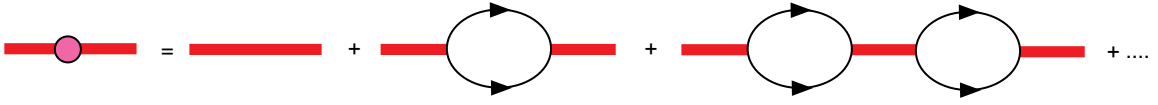}} }
\vskip 0.15in
\noindent
\caption{\it The dressed \tran{} propagator. 
The bare \tran{} propagator
is dressed by nucleon bubbles to all orders. 
Each diagram
counts as $Q^{-2}$ in the power-counting scheme.
}
\label{fig:trans}
\vskip .2in
\end{figure}
It is easy to show that this Lagrange density alone reproduces 
the NN scattering amplitude when
\begin{eqnarray}
\yt^2 & = & {8\pi\over M_N^2 r^{(\siii)}}
\ \ ,\ \ 
\Delta\ =\ {2\over M_N  r^{(\siii)}} \left({1\over a^{(\siii)}} - \mu\right)
\ \ \ ,
\label{eq:coupfix}
\end{eqnarray}
where $\mu$ is the renormalization scale. As far as power-counting is
concerned $\yt \sim \sqrt{Q}$ and $\Delta \sim Q^2$. Therefore, the
bare \tran{} propagator counts as $Q^{-2}$, as do arbitrary insertions
of nucleon bubbles. Hence the bubbles must be summed to all orders as
in Fig.~\ref{fig:trans}.  The \tran{} propagator dressed with
nucleon bubbles, $D^{(\siii)} (\overline{E})$, as a function of
center-of-mass energy $\overline{E}$, is
\begin{eqnarray}
D^{(\siii)} (\overline{E}) & = & 
{4\pi\over M_N \yt^2}
\ {i\over 
\mu + {4\pi\over M_N \yt^2}\Delta
- {4\pi\over M_N \yt^2} \overline{E}
+ i \sqrt{M_N \overline{E}}}.
\label{eq:tprop}
\end{eqnarray}
The LO scattering amplitude is then
\begin{eqnarray}
{\cal A}_{-1}& = & 
{{4\pi}\over M_N}
{1\over -{1\over a^{(\siii)}} + {1\over 2} r^{(\siii)} M_N \overline{E}
- i \sqrt{M_N \overline{E}} }
\ \ \ ,
\label{eq:amplLO}
\end{eqnarray}
which is simply the effective range expansion, 
neglecting the shape parameter and higher contributions.

\begin{figure}[!ht]
\centerline{{\epsfxsize=4.5in \epsfbox{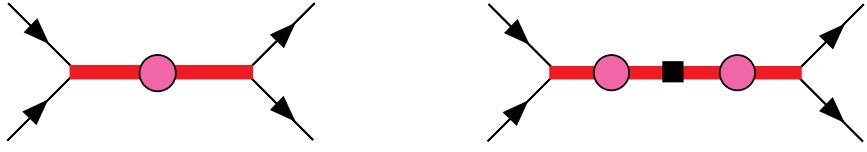}} }
\vskip 0.15in
\noindent
\caption{\it 
Feynman diagrams that contribute to NN
scattering.
The thick solid lines denote the \tran{} field, 
while the single lines
denote the nucleon field. 
The black box is an insertion of the
shape parameter correction that is suppressed by $Q^3$
compared to the LO diagram.
}
\label{fig:shape}
\vskip .2in
\end{figure}

To satisfy oneself that the Lagrange density in eq.~(\ref{eq:trandef})
reproduces the scattering amplitude that one would obtain by writing
down all possible four-nucleon operators, as is done in $\nopi$, it is
sufficient to compare the scattering amplitude in the two theories.
The exchange of a fully-dressed \tran{} between nucleons is identical
to the exact bubble sum, up to terms beyond the effective
range,\footnote{Defined via ${\bf p}\cot\delta = -{1\over a} + {1\over
2} r |{\bf p}|^2 + r_1 |{\bf p}|^4+...$} i.e. the shape parameter
$r_1^{(\siii)}$, and higher.  To include the shape parameter, for
instance, one includes a term in the Lagrange density,
\begin{eqnarray}
{\cal L}^{\rm shape} & = & 
-{2 M_N r_1^{(\siii)}\over r^{(\siii)}}\ 
\left[ {\cal O}_E t_j\right]^\dagger
\left[ {\cal O}_E t^j\right]
\ \ \ .
\label{eq:shape}
\end{eqnarray}
This operator is suppressed by $Q^3$ relative to $\Delta$ and
therefore gives rise to a perturbative correction to the LO amplitude
via the right diagram in Fig.~\ref{fig:shape}. 
The $Q^3$ suppressed  amplitude is
thus
\begin{eqnarray}
{\cal A}_{2}& = & 
-{{4\pi}\over M_N} r_1^{(\siii)} ({M_N}\overline{E})^2
{1\over\left( -{1\over a^{(\siii)}} + {1\over 2} r^{(\siii)} M_N \overline{E}
- i \sqrt{M_N \overline{E}}\right)^2 }
\ \ \ .
\label{eq:amplNLO}
\end{eqnarray}
%

The Lagrange density in eq.~(\ref{eq:shape}) demonstrates a general feature 
of transforming between a theory written in terms of only 
four-nucleon operators and one written in terms of \trans.
The ``rule'' for replacing a nucleon bi-linear with a \tran{} field is,
up to numerical factors,
\begin{eqnarray}
N^T P^j N &\rightarrow &
{1\over \sqrt{ M_N r^{(\siii)} }}\ t^j
\ \ \ ,\ \ \ 
N^T \overline{P}^a N \rightarrow 
{1\over \sqrt{ M_N r^{(\si)} }}\ s^a
\ \ \ ,
\label{eq:rule}
\end{eqnarray}
where $\overline{P}^a$ is the projector for the $\si$ channel and 
$s^a$ is the $\si$ \tran.
This is important to keep in mind as it introduces factors of $\sqrt{Q}$
into the coefficients of operators. The power counting with the effective
range scaling as $1/Q$ and a Lagrange density written in terms of \tran{} fields
will be denoted by $\nopit$.

In the most general Lagrange density that describes interactions in or
with the $\si$ and $\siii$ channels, all higher dimension operators 
involve couplings to 
the \tran{} fields, and not to the nucleon bilinear $N^TP^jN$ in the 
$\siii$ channel
or $N^T P^aN$ in the $\si$ channel.
The simplest example of this is perhaps the $\siii-\diii$ channels.
The Lagrange density describing mixing is, up to NLO
\begin{eqnarray}
{\cal L}^{(sd)} & = & 
-{ C_2^{(sd)}\over \sqrt{ M_N r^{(\siii)} }}\ 
t^\dagger_j
\ N^T {\cal O}^{ij}_{\bf (2)}\   P^i N
\ -\ {C_4^{(sd)}\over \sqrt{ M_N r^{(\siii)} }}\ 
\left[ {\cal O}_E t_j\right]^\dagger 
\ N^T {\cal O}^{ij}_{\bf (2)}\   P^i N
\ \ +\ \  {\rm h.c.}
\ \ ,
\label{eq:lagsd}
\end{eqnarray}
with
\begin{eqnarray}
{\cal O}^{ij}_{\bf (2)} 
& = & -{1\over 4}\left[
(\overleftarrow\nabla-\overrightarrow\nabla)^i 
(\overleftarrow\nabla-\overrightarrow\nabla)^j 
-{1\over n-1}\delta^{ij} (\overleftarrow\nabla-\overrightarrow\nabla)^2
\right]
\ \ \ ,
\label{eq:dop}
\end{eqnarray}
and the ellipses denote operators involving more powers of the center-of-mass
energy $\overline{E}$.
This leads to a mixing parameter $\overline{\varepsilon}_1$ at NLO of
\begin{eqnarray}
\overline{\varepsilon}_1 & = & 
\left[ C_2^{(sd)}\ +\  C_4^{(sd)} \overline{E}\right]
{ \sqrt{M_N} (M_N \overline{E})^{3/2}\over 6\sqrt{\pi}}
{1\over \sqrt{ ( -{1\over a^{(\siii)}} 
+ {1\over 2} r^{(\siii)} M_N \overline{E})^2 
+ M_N \overline{E}}}
\ \ \ .
\label{eq:epone}
\end{eqnarray}
Writing this expression in terms of the asymptotic $\siii$ to $\diii$ ratio
$\eta_{sd}$, defined by 
\begin{eqnarray}
\etasd & = & -\tan\left(\varepsilon_1\right)
\ \ \ ,\ \ \ 
\tan\left(2 \varepsilon_1\right)
\ = \ 
{ \tan\left(2 \overline{\varepsilon}_1\right)
\over
\sin\left(\overline{\delta}_0-\overline{\delta}_2\right)}
\ \ \ ,
\label{eq:asymDS}
\end{eqnarray}
evaluated at the deuteron pole, $|{\bf p}|=i\gamma$,
\begin{eqnarray}
\overline{\varepsilon}_1 & = & 
\eta_{sd} \left({M_N\overline{E}\over\gamma^2}\right)
{\sqrt{M_N\overline{E}}\over \sqrt{ ( -{1\over a^{(\siii)}} 
+ {1\over 2} r^{(\siii)} M_N \overline{E})^2 
+ M_N \overline{E}}}
\ +\ ...
\ \ \ .
\label{eq:eponeB}
\end{eqnarray}
The ellipses denote the contribution from the 
$C_4^{(sd)}$ and higher operators.

\section{External Probes}

For processes involving the deuteron either in the initial, or final states, 
or both, 
it is convenient to use sources that couple to the \tran{} fields directly.
As we only require the source to have a non-zero overlap with the state of
interest, sources coupling to the \tran{} fields are sufficient.
In fact, this makes calculations much simpler.
To compute S-matrix elements, the residue of the $\siii$ 
\tran{} field is required at the deuteron pole.
Writing the \tran{} propagator as
\begin{eqnarray}
G (\overline{E}) & = & {i {\cal Z}(\overline{E})\over \overline{E}+B}
\ =\ 
{i ( Z_d + Z_1 (\overline{E}+B)  + ...) \over \overline{E}+B}
\ \ \ ,
\label{tpropexp}
\end{eqnarray}
gives
\begin{eqnarray}
Z_d & = & {\gamma r^{(\siii)}\over 1-\gamma  r^{(\siii)}}
\ \ \ .
\label{eq:Zdef}
\end{eqnarray}
In order to demonstrate the simplifications that arise for processes
involving electroweak currents, we consider the deuteron charge form factor,
$np\rightarrow d\gamma$ and Compton scattering on the deuteron.

\subsection{Deuteron Electric Form Factor}

\begin{figure}[!ht]
\centerline{{\epsfxsize=4.5in \epsfbox{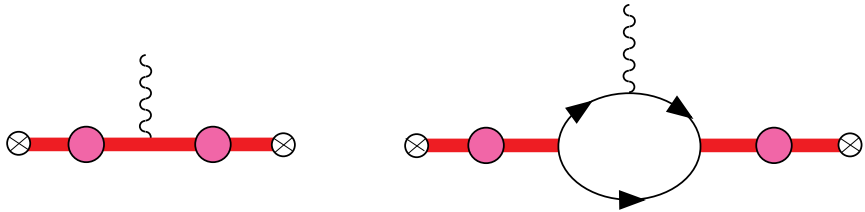}} }
\vskip 0.15in
\noindent
\caption{\it
The LO Feynman diagrams that contribute to the deuteron
charge form factor.
The thick solid lines denote the \tran{} field, 
while the thin lines
with arrows  denote the nucleon field.
The crosses are sources of the \tran{} field.
LSZ reduction removes the fully dressed \tran{} propagators 
from the Green functions.
}
\label{fig:CFF}
\vskip .2in
\end{figure}

The LO diagrams that contribute to the deuteron charge form factor in $\nopit$
are shown in Fig.~\ref{fig:CFF}.
%
In addition to diagrams where the photon couples to the nucleon, there are
also couplings to the \tran{} field obtained by gauging the Lagrange density
in eq.~(\ref{eq:trandef}).
At higher orders there are contributions from higher dimension operators
involving more derivatives on the nucleon field, such as the nucleon charge 
radius operator, and also from higher dimension couplings of the 
\tran, such as the \tran{} charge radius.
It is interesting to note that these higher dimension operators are not
renormalized by the s-wave 
strong interactions, as is the case when four-nucleon
operators alone are used to describe the s-wave scattering amplitude.
Therefore, the size of the contribution of the higher dimension operators
can be estimated simply by counting the dimensions of the operator and
including the appropriate powers of the high scale $\Lambda$ in its
coefficient.
The contribution to the charge form factor 
from the tree-level photon coupling to the \tran{}
is, (removing a factor of $i e$ from the amplitude)
\begin{eqnarray}
\Gamma^{(a)} & = & 1\  \left(\sqrt{Z_d}\right)^2
\ \ \ ,
\label{eq:CFFa}
\end{eqnarray}
and from the nucleon loop diagram is
\begin{eqnarray}
\Gamma^{(b)} & = & -{4\over |{\bf q}|\  r^{(\siii)}}
\tan^{-1}\left({ |{\bf q}|\over 4\gamma}\right)
 \left(\sqrt{Z_d}\right)^2
\ \ \ ,
\label{eq:CFFb}
\end{eqnarray}
where $|{\bf q}|$ is the magnitude of the photon three-momentum.
These contributions combine to give a charge form factor of
\begin{eqnarray}
F_C ( |{\bf q}| )
& = & 
{\gamma r^{(\siii)}\over 1-\gamma\ r^{(\siii)}}
\left[{4\over |{\bf q}|\  r^{(\siii)}}
\tan^{-1}\left({ |{\bf q}|\over 4\gamma}\right)\ -\ 1\ \right]\ \ \ .
\label{eq:CFFtot}
\end{eqnarray}
This reproduces the result of ER theory, as expected.
At higher orders in the effective field theory expansion there
will be a contribution from the nucleon form factor, starting with the 
nucleon charge radius\cite{Ch99} at order $Q^2$.
At order $Q^3$ there is a contribution from the  \tran{} charge radius
operator.

\subsection{$np\rightarrow d\gamma$}

The radiative capture process $np\rightarrow d\gamma$
at low-energy has been computed with $\sim 1\%$
accuracy with $\nopi$~\cite{CS99,Ru99}.
The process is dominated by the $E1$ and $M1$ multipoles, for
which the matrix element is
\begin{eqnarray}
{\cal A} & = & 
e\  X_{E1}\ U^T_n \sigma_2\  \sigma\cdot\epsilon_{(d)}^* U_p
\ {\bf p}\cdot\epsilon_{(\gamma)}^* 
\ +\ 
i\  e \  X_{M1} \ \varepsilon^{abc}\ \epsilon_{(d)}^{* a}\  {\bf k}^b\
\epsilon_{(\gamma)}^{* c} \ U^T_n \sigma_2 U_p
\ \ ,
\label{eq:nprate}
\end{eqnarray}
where ${\bf k}$ and $\epsilon_{(\gamma)}$ are the photon momentum and 
polarization vector,
$\epsilon_{(d)}$ is the deuteron polarization vector and 
$U_{n,p}$ are the neutron and proton spinors;
${\bf p}$ is the momentum of the proton in the center-of-momentum frame.
It is convenient to define dimensionless variables~$\tilde{X}_{\pi L}$,
\begin{eqnarray}
{ |{\bf p}| M_{N}\over \gamma^{2} + |{\bf p}|^2}
X_{E1}\  & = &
i {2\over M_N} \sqrt{\pi\over\gamma^3}\tilde{X}_{E1}
\quad ,\quad
X_{M1}\  \ =\ 
i {2\over M_N} \sqrt{\pi\over\gamma^3}\tilde{X}_{M1}
\ \ \ ,
\label{eq:tildedef}
\end{eqnarray}
in terms of which the total cross section is
\begin{eqnarray}
\sigma & = & {4\pi \alpha (\gamma ^{2}+|{\bf p}|^2)^3 \over M_{N}^{4} \gamma ^{3} |{\bf p}|}
\left[ \ 
|\tilde{X}_{M1}|^{2}  \ +\ 
|\tilde{X}_{E1}|^{2}\ 
\right]
\ \ \ .
\label{eq:unpol}
\end{eqnarray}

%
\begin{figure}[!ht]
\centerline{{\epsfxsize=3.5in \epsfbox{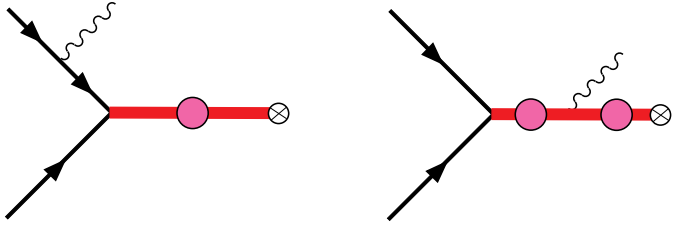}} }
\vskip 0.15in
\noindent
\caption{\it The diagrams providing the LO
contribution to the $E1$ amplitude, $ \tilde X_{E1}$.
The photon is minimally coupled.
}
\label{fig:Eone}
\vskip .2in
\end{figure}
The diagrams shown in Fig.~\ref{fig:Eone} provide the 
LO contributions to the $E1$ amplitude in $\nopit$.
Up to order $Q^2$, the $E1$ amplitude is given by 
\begin{eqnarray}
\tilde X_{E1} & = & -{1\over\sqrt{1-\gamma r^{(\siii)}}}\ 
{|{\bf p}| M_N \gamma^2 \over (|{\bf p}|^2+\gamma^2)^2}
\ \ \ ,
\label{eq:Eoneamp}
\end{eqnarray}
consistent with ER theory at this order.
A higher dimension operator coupling an E1 photon to a $\siii$-\tran{}
and a p-wave two nucleon final state is suppressed by $Q^3$ in the power
counting.

The $M1$ amplitude receives contributions from the magnetic moments
of the nucleon and from four-nucleon operators coupling to the magnetic field,
which are described by the Lagrange density involving \tran{} fields
\begin{eqnarray}
{\cal L}^{\bf B} & = & 
{e\over 2 M_N} N^\dagger \left( \kappa_0+\kappa_1 \tau^3\right)
\sigma\cdot {\bf B}\  N
\ +\ 
e\ { L_1\over M_N \sqrt{ r^{(\si)}\ r^{(\siii)}}}
\ t^{j\dagger} s_3 {\bf B}_j
\ \ \ +\ \  {\rm h.c.},
\end{eqnarray}
where $t_j$ is the $\siii$ \tran{} and 
$s_a$ is the $\si$ \tran.
The unknown coefficient $L_1$, which contributes at order $Q$
(compared to the $\kappa_1$ contributions)
must either be predicted from
QCD or determined experimentally in order to have 
model-independent predictive power.
%
\begin{figure}[!ht]
\centerline{{\epsfxsize=3.5in \epsfbox{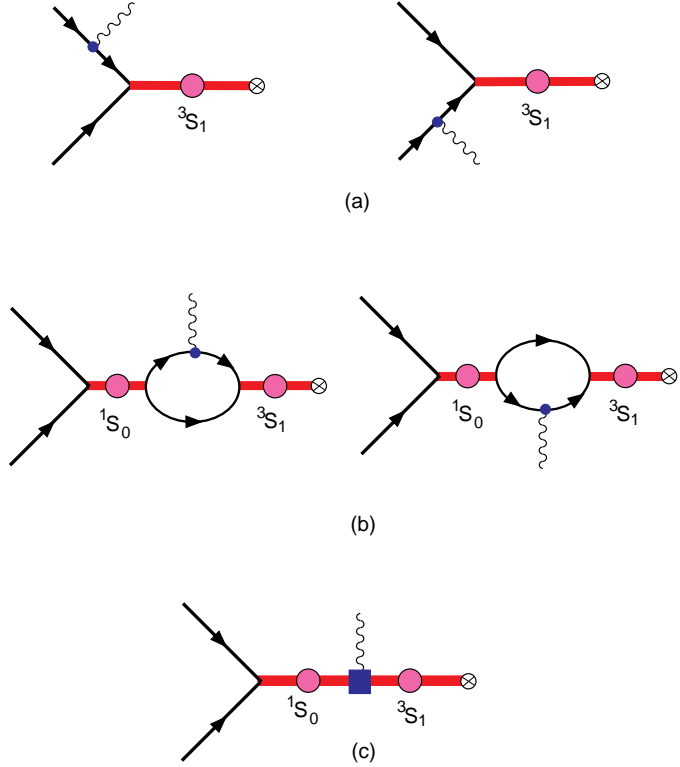}} }
\vskip 0.15in
\noindent
\caption{\it 
Diagrams contributing to the $M1$ amplitude for 
$np\rightarrow d\gamma$
capture from the $\si$ channel.
The thick lines correspond to either $\si$ or $\siii$ 
\trans,
while the thin lines represent nucleons.
The solid circles correspond to an insertion of the 
single nucleon $\sigma\cdot {\bf B}$ operator.
The solid square in diagram (c) denotes the insertion of a 
four-nucleon-magnetic-photon operator described by a 
coupling between the $\si$-\tran{} and the $\siii$-\tran{} and a magnetic
photon.
}
\label{fig:Mone}
\vskip .2in
\end{figure}
The diagrams in Fig.~\ref{fig:Mone} (a) and (b) are the LO contributions
to the $M1$ amplitude for capture from the $\si$ channel.
The diagram in Fig.~\ref{fig:Mone} (c) enters at NLO and represents the 
contribution from the four-nucleon-one-magnetic-photon operator at this order.
It is straightforward to show that the sum of the diagrams in 
Fig.~\ref{fig:Mone} gives
\begin{eqnarray}
\tilde X_{M1} & = & 
{1\over\sqrt{1-\gamma r^{(\siii)} }}
\ 
{1\over  -{1\over a^{(\si)}} + {1\over 2} r^{(\si)} |{\bf p}|^2 - i|{\bf p}|}
\nonumber\\
& & \left[\ 
\kappa_1 \ {\gamma^2\over |{\bf p}|^2 + \gamma^2} 
\left( \gamma - {1\over a^{(\si)}} + {1\over 2} r^{(\si)} |{\bf p}|^2 \right)
\ +\ 
L_1\  {\gamma^2\over 2} \  
\right]
\ \ \ ,
\label{eq:xmonet}
\end{eqnarray}
which reproduces the results of Ref.~\cite{CS99} when expanded out to 
lowest order in the effective ranges.
In calculating $\tilde X_{M1}$ in eq.~(\ref{eq:xmonet}) we have used
non-relativistic kinematics, as the relativistic corrections are small
as shown in Ref.~\cite{Ru99}.
In order to reproduce the measured cross section of 
$\sigma^{\rm expt} = 334.2\pm 0.5~{\rm mb}$\cite{CWCa}
for an incident neutron speed of $|{\bf v}|=2200~{\rm m/s}$ in the 
laboratory frame,  $L_1$ is required to be
$L_1 \sim -4.0~{\rm fm}$ (where we have worked to linear order in $L_1$)

The numerical values of the cross section for $np\rightarrow d\gamma$
one obtains from the expressions in eq.~(\ref{eq:Eoneamp}) and
eq.~(\ref{eq:xmonet}) agree very well with the values obtained in
Refs.~\cite{CS99,Ru99}, as expected.  The uncertainty in these
expressions translates into an uncertainty in the capture cross
section in the energy regime relevant for nucleosynthesis of $\sim
3\%$.  The difference between this work and that of
Refs.~\cite{CS99,Ru99}, is the complexity of the computation.  In
Refs.~\cite{CS99,Ru99} multiple (unnested) loop diagrams were computed
and the renormalization group evolution of $L_1 (\mu)$ was
non-trivial.  In the present computation only one-loop diagrams occur,
with the majority of the calculation being tree-level.  In addition,
$L_1$ is not renormalized by the s-wave interactions at this order.

\subsection{$\gamma d\rightarrow \gamma d$ Compton Scattering}

Finally, we examine Compton scattering in $\nopit$.  Working in the
zero-recoil limit, and neglecting relativistic effects (both of which
are very small) we compute the cross section for the scattering of
unpolarized photons from unpolarized deuterons in the low energy
regime.  This process has been computed in $\nopi$ in Ref.~\cite{RG00}.
Writing the amplitude for this process in terms of two form factors,
$T_1$ and $T_2$, we have
\begin{eqnarray}
{\cal A} & = & 
i {e^2\over 2 M_N} 
\left[\ 
T_1\ \epsilon\cdot\epsilon^{\prime *}
\ +\ T_2\ 
\left(\epsilon\times {\bf k}\right)\cdot 
\left(\epsilon^{\prime *}\times {\bf k}^\prime\right)
\right]
\ \epsilon_{(d)}\cdot  \epsilon_{(d)}^{\prime *}
\ \ \ ,
\label{eq:gdamp}
\end{eqnarray}
where ${\bf k}$ and $\epsilon$ are the three-momenta and polarization vector
of the incident photon, and 
${\bf k}^\prime$ and $\epsilon^\prime$ 
are the three-momenta and polarization vector
of the scattered photon;
$\epsilon_{(d)}$ and  $\epsilon_{(d)}^\prime$ are the polarization vectors
of the initial and final state deuteron.
As we are neglecting recoil and relativistic effects we will not recover 
the factor of $M_d$ in the coefficient of eq.~(\ref{eq:gdamp}).
It was shown in Ref.\cite{Ch99} that $M_d$ is recovered order-by-order
in perturbation theory in $1/M_N$.
%
\begin{figure}[!ht]
\centerline{{\epsfxsize=3.5in \epsfbox{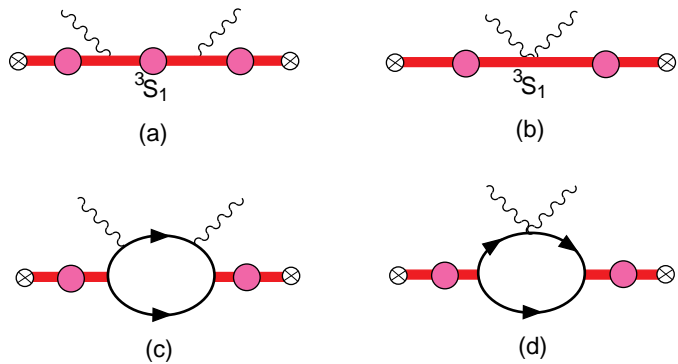}} }
\vskip 0.15in
\noindent
\caption{\it 
Leading-order diagrams contributing to the $T_1$ amplitude for 
$\gamma d\rightarrow \gamma d$ Compton scattering.
The thick lines correspond to a $\siii$ \tran,
while the thin lines represent nucleons.
The photons are minimally coupled to both the nucleon and 
the \tran.
Crossed diagrams are not shown.
}
\label{fig:Elec}
\vskip .2in
\end{figure}
The differential cross section resulting from eq.~(\ref{eq:gdamp}) is
\begin{eqnarray}
{d\sigma\over d\Omega_{\rm lab}}
& = & 
{\alpha^2\over 8 M_N^2}
\left[\ 
\left( |T_1|^2+|T_2|^2\right)\left(1+\cos^2\theta\right)
\ +\ 
4 {\rm Re}\left( T_1 T_2^*\right)\cos\theta
\ \right]
\ \ \ .
\label{eq:diff}
\end{eqnarray}

\begin{figure}[!ht]
\centerline{{\epsfxsize=1.5in \epsfbox{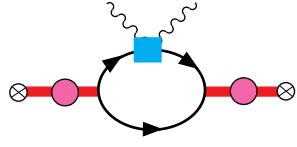}} }
\vskip 0.15in
\noindent
\caption{\it 
Contribution to $T_1$ and $T_2$ from the electric and magnetic
polarizabilities of the nucleon.
The solid square denotes an insertion of the 
nucleon polarizability operators.
}
\label{fig:Epol}
\vskip .2in
\end{figure}

For the purposes of this discussion we will use 
Regime II Q-counting~\cite{CGSSb} where $\gamma\sim\omega\sim Q$
to determine the size of a 
diagram contributing to the scattering amplitudes.
The diagrams shown in Fig.~\ref{fig:Elec} 
contribute to the electric form factor $T_1$ in eq.~(\ref{eq:diff}). 
Figs.~\ref{fig:Elec} (a), (b)
and (d) contribute at order $Q^0$.
Fig.~\ref{fig:Elec} (c) contributes at order $Q^{1/2}$ and higher.
It is straightforward to show that
\begin{eqnarray}
T_1^{+\omega} & = & 
{1\over 1-\gamma  r^{(\siii)}}
\left[\ 
{1\over 2} \gamma r^{(\siii)}
\ +\ 
{2\gamma \over 3}{\gamma + 2\sqrt{\gamma^2 + M_N\omega - i \epsilon}\over
\left(\gamma + \sqrt{\gamma^2 + M_N\omega - i \epsilon}\right)^2}
\right.
\nonumber\\
&  &  \left.
\qquad 
-\ 
{4\gamma \over \omega \sqrt{2-2\cos\theta}}
\tan^{-1}\left({\omega \sqrt{2-2\cos\theta}\over 4\gamma}\right)
\ +\ 
{\omega^2\cos\theta\over 6}
{\gamma + 3\sqrt{\gamma^2 + M_N\omega - i \epsilon}\over
\left(\gamma + \sqrt{\gamma^2 + M_N\omega - i \epsilon}\right)^3}
\right.
\nonumber\\
&  &  \left.
\qquad 
\ +\ 
{4 M_N\pi \left(\alpha_p+\alpha_n\right)\omega^2\over e^2}
{4\gamma \over \omega \sqrt{2-2\cos\theta}}
\tan^{-1}\left({\omega \sqrt{2-2\cos\theta}\over 4\gamma}\right)
\ \right]
\ \ \ ,
\end{eqnarray}
where $\omega$ is the incident photon energy.
The electric form factor is
$T_1=T_1^{+\omega}+T_1^{-\omega}$,
with $T_1\rightarrow -1$ in the $\omega\rightarrow 0$ limit,
as required.
We have included the contribution from the nucleon electric 
polarizabilities, which are defined in the Lagrange density
\begin{eqnarray}
{\cal L}_1 & = &
2\pi \alpha_n \ n^\dagger n \ {\bf E}^2
\ +\ 
2\pi \alpha_p \ p^\dagger p \ {\bf E}^2
\ +\ 
2\pi \beta_n \ n^\dagger n \ {\bf B}^2
\ +\ 
2\pi \beta_p \ p^\dagger p \ {\bf B}^2
\ \ \ ,
\label{eq:lagpol}
\end{eqnarray}
and which contribute through the diagrams shown in  Fig.~\ref{fig:Epol}
at order $Q^2$.
%
The zero-range limit of $T_1$ coincides with previous 
computations~\cite{CGSSb,LLa}.

%
\begin{figure}[!ht]
\centerline{{\epsfxsize=4.0in \epsfbox{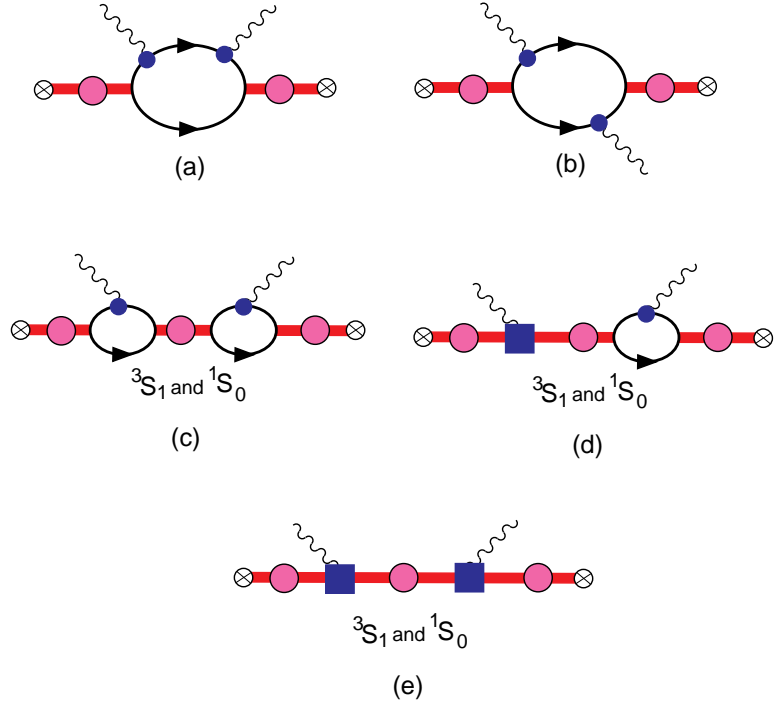}} }
\vskip 0.15in
\noindent
\caption{\it 
Diagrams contributing to the $T_2$ amplitude for 
Compton scattering.
The thick lines correspond to either 
a $\si$ or $\siii$ \tran,
while the thin lines represent nucleons.
The photons couple to the nucleons via the 
$\sigma\cdot {\bf B}$
interaction, represented by the small solid circle.
The solid square denotes coupling between the photon and \trans{}
due to the higher dimension operators, e.g. the $L_1$ operator
that contributes to $np\rightarrow d\gamma$.
Crossed diagrams are not shown.
}
\label{fig:Mag}
\vskip .2in
\end{figure}
The magnetic form factor is dominated by the diagrams shown in 
Fig.~\ref{fig:Mag}, and can be shown to be
\begin{eqnarray}
T_2^{+\omega} & = & 
{1\over 1-\gamma r^{(\siii)}}
\left[\ 
{4\kappa_1^2\gamma\over 3 M_N^2\omega^2}
{ \left(\gamma-\sqrt{\gamma^2+M_N\omega- i \epsilon}\right)
\left(\gamma-{1\over a^{(\si)}} 
- {1\over 2}r^{(\si)}\left(\gamma^2+M_N\omega\right)\right)\over
-{1\over a^{(\si)}} 
- {1\over 2}r^{(\si)}\left(\gamma^2+M_N\omega\right)
+\sqrt{\gamma^2+M_N\omega- i \epsilon}}
\right.\nonumber\\
& & \left.
\quad
+\ 
{4\gamma\kappa_1 L_1 \over 3}
{1\over \gamma + \sqrt{\gamma^2 + M_N\omega - i \epsilon}}
{1\over -{1\over a^{(\si)}} 
- {1\over 2}r^{(\si)}\left(\gamma^2+M_N\omega\right)
+\sqrt{\gamma^2+M_N\omega- i \epsilon}}
\right.\nonumber\\
& & \left.
\quad
+\ 
{\gamma L_1^2\over 3}
{1\over -{1\over a^{(\si)}} 
- {1\over 2}r^{(\si)}\left(\gamma^2+M_N\omega\right)
+\sqrt{\gamma^2+M_N\omega- i \epsilon}}
\right.\nonumber\\
& & \left.
\quad
+\ 
{2\kappa_0^2\gamma (r^{(\si)})^2\over 3}
{1\over \sqrt{\gamma^2+M_N\omega- i \epsilon} 
- \gamma - {1\over 2}M_N r^{(\si)}\omega}
\right.\nonumber\\
& & \left.
\quad
-\ 
{1\over 6} 
{\gamma + 3\sqrt{\gamma^2 + M_N\omega - i \epsilon}\over
\left(\gamma + \sqrt{\gamma^2 + M_N\omega - i \epsilon}\right)^3}
\right.\nonumber\\
& & \left.
\quad
\ +\ 
{4 M_N\pi \left(\beta_p+\beta_n\right)\over e^2}
{4\gamma \over \omega \sqrt{2-2\cos\theta}}
\tan^{-1}\left({\omega \sqrt{2-2\cos\theta}\over 4\gamma}\right)
\right]
\ \ \ ,
\label{eq:Mag}
\end{eqnarray}
where $T_2=T_2^{+\omega}+T_2^{-\omega}$ and where we have included the
contribution from the magnetic polarizabilities of the
nucleon, $\beta_p$ and $\beta_n$, entering from diagrams of the form
shown in Fig.~\ref{fig:Epol}. Figs.~\ref{fig:Mag}(a-c) are the leading
contributions at order $Q$ and higher, 
Fig.~\ref{fig:Mag}(d) contributes at order $Q^2$ and higher while
Fig.~\ref{fig:Mag}(e) contributes at order $Q^3$.
The nucleon magnetic polarizabilities enter at order $Q^2$.

\begin{figure}[!ht]
\centerline{{\epsfxsize=2.0in \epsfbox{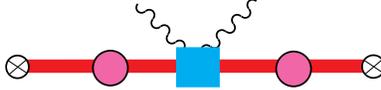}} }
\vskip 0.15in
\noindent
\caption{\it 
Insertion of a higher dimension operator contributing to
Compton scattering.
The solid square denotes an insertion of the \tran{} electric
or magnetic polarizability.
}
\label{fig:Higher}
\vskip .2in
\end{figure}

It is clear from these expressions that they are precisely what one would
obtain from ER theory when only the  single nucleon operators are retained.
However, deviations from ER theory occur due to the presence of four-nucleon
local interactions with the electric and 
magnetic field\cite{CGSSb,CGSSa}, 
such as the $L_1$ operator and also
interactions of the form
\begin{eqnarray}
{\cal L} & = & 
{2\pi\alpha_4\over M_N r^{(\siii)}} 
t_j^\dagger t^j |{\bf E}|^2 
\ +\ 
{2\pi\beta_4\over M_N r^{(\siii)}} 
t_j^\dagger t^j |{\bf B}|^2 
\ +\ ...\ \ ,
\label{eq:fourpol}
\end{eqnarray}
that directly couple \trans{} to two-photons, entering at the same
order as the $L_1^2$ contribution in eq.~(\ref{eq:Mag}).  The
contributions to Compton scattering from these higher dimension
operators, as shown in Fig.~\ref{fig:Higher}, are of order $Q^3$, and
so enter at one order higher than the nucleon polarizabilities
themselves.  So it appears that the nucleon polarizabilities can be
determined from Compton scattering only with precision of order $Q$,
due to the presence of higher dimension operators, corresponding to
the polarizabilities of the \tran{} itself.
%

\begin{figure}[!ht]
\centerline{{\epsfxsize=3.0in \epsfbox{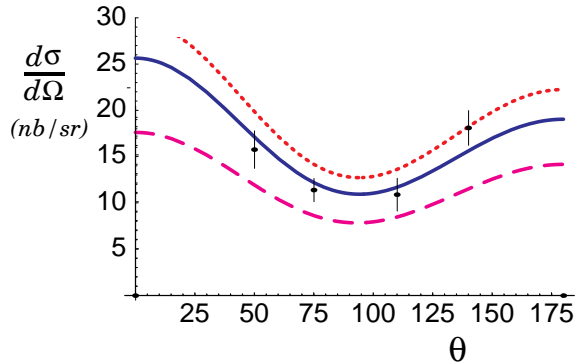}} }
\vskip 0.15in
\noindent
\caption{\it 
Differential cross section for 
$\gamma d\rightarrow \gamma d$
for an incident photon energy of $\omega=49~{\rm MeV}$.
The dashed curve is the zero-range limit of the cross section,
i.e. $r^{(\si)}=r^{(\siii)}=0$.
The dotted curve corresponds to the full calculation but with
the nucleon polarizabilities set equal to zero, i.e. 
$\alpha_N=\beta_N=0$.
The solid curve is the full calculation including the 
nucleon polarizabilities being set equal to their value 
computed at one-loop in
chiral perturbation theory~\protect\cite{BKMa}.
The data is from Ref.~\protect\cite{Lucas}.
}
\label{fig:cross}
\vskip .2in
\end{figure}

The differential cross section for an incident photon energy of
$\omega=49~{\rm MeV}$
resulting from the calculated $T_1$ and $T_2$
is shown in Fig.~\ref{fig:cross}.
Even though $\omega=49~{\rm MeV}$ is quite possibly above the
energy at which the pionless theory makes sense, it is nonetheless
useful to see the impact that the nucleon polarizabilities have
on the differential cross section.
The solid curve in  Fig.~\ref{fig:cross} includes the contribution
from the individual nucleon polarizabilities
as computed at one-loop order in chiral perturbation theory~\cite{BKMa},
\begin{eqnarray}
  \alpha_N & = & 10\beta_N\ =\ 
{5 g_A^2 e^2\over 192\pi^2 f_\pi^2 m_\pi}\ =\ 
1.2\times 10^{-3}\ {\rm fm}^3
\ \ \ \ .
\label{eq:pols}
\end{eqnarray}
It is clear that high precision measurements of the Compton scattering
cross section at low-energies will allow for a reasonable 
determination of the nucleon isoscalar electric polarizability,
and possibly the isoscalar magnetic polarizability.
As the contribution from the nucleon polarizabilities scales like 
$\omega^2$, the lower the energy of the incident photon the higher the 
precision of the measurement that is required to provide
a comparable constraint on the polarizabilities.
%
%

\section{Conclusion}

We have solidified three ideas that allow for much simpler
calculations of low-energy nuclear observables using effective field
theory. First, we have shown that by using the nucleon equations of
motion and judiciously integrating by parts, the number of higher
dimension four-nucleon operators is greatly reduced. Second, we have
suggested that the large corrections arising from the effective range
parameter in both the $\si$ and $\siii$ channels can be resummed by
taking the range parameter to scale as $1/Q$, just like the scattering
length. Third, we have taken the idea of using \tran{} fields
seriously and have shown that their use reduces the labor of computing
higher order corrections in processes with external gauge fields
substantially. Unlike in $\nopi$, in $\nopit$ two-nucleon higher
dimension operators are not renormalized by leading-order four-nucleon
operators, and hence one can estimate the contributions of such
operators by naive dimensional analysis, in terms of the high scale
$\Lambda$. It is important to realize that $\nopit$ does not
invalidate $\nopi$ but rather allows the EFT to accomodate a large
effective range.

As the theory written in terms of \tran{} fields behaves very
differently under renormalization scale evolution than the theory
written in terms of local four-nucleon operators, it is conceivable
that this new power counting leads to simplifications in the
three-nucleon systems beyond those utilized in
Ref.~\cite{bosons,harald,haraldb}. A step in this direction has
already been taken in Ref.~\cite{ussr}. This possibility will be
further explored in the near future.

It may be the case that this approach will tame the convergence
problems encountered in the pionful theory~\cite{FMSa} with KSW power
counting~\cite{Ka98B,Ka98A,Ka99C}.  
In KSW counting the leading order phase
shift in both the $\si$ and $\siii$ channels tends to $\pi/2$, at
energies below $m_\pi/2$ (the scale near which the perturbative theory
appears to diverge).  However, with the \tran{} describing the leading
order amplitude, the phase shift is much smaller in this same energy
region, and consequently the contribution from higher orders in the
KSW expansion are expected to be suppressed by factors of two or
three.  This idea is yet to be explored, and it will be interesting to
see whether this scheme can be made consistent with a perturbative
pion.

\vskip 0.5in

We would like to thank David Kaplan for useful discussions. We
acknowledge useful and in some cases entertaining correspondence with
Paulo Bedaque, Jiunn-Wei Chen, Harald Grie\ss hammer, Hans-Werner
Hammer, Gautam Rupak, Roxanne Springer and Ubi van Kolck.  This work
is supported in part by the U.S. Department of Energy under Grant No.
DE-FG03-97ER4014.

\end{document}